# Investigation of Microstructural Evolution in All-Solid-State Micro-Batteries through *in situ* Electrochemical TEM

*Sorina Cretu[1,2], Nicolas Folastre[2,3], David Troadec[5], Ingrid Marie Andersen[1],*
*Rainer Straubinge[6], Nynke A. Krans[6], Stéphane Aguy[6], Arash Jamali[2,3],*
*Martial Duchamp[1], Arnaud Demortière[2,3,4]\**

[1] School of Materials Science and Engineering, Nanyang Technological University, 50 Nanyang Avenue, 639798, Singapore

[2] Laboratoire de Réactivité et de Chimie des solides (LRCS), Université de Picardie Jules Verne, CNRS UMR 7314, 33 rue Saint Leu, 80039 Amiens Cedex, France

[3] Réseau sur le stockage Electrochimique de l'Energie (RS2E), CNRS FR 3459, 33 rue Saint Leu, 80039 Amiens Cedex, France

[4] ALISTORE-ERI, FR CNRS, 80039, 3104, France

[5] Institut d'Electronique, de Microélectronique et de Nanotechnologie, Université de Lille, France

[6] Protochips, 3800 Gateway Centre Blvd, 306, Morrisville, North Carolina 27560, USA

**Corresponding author:** Arnaud.demortiere@cnrs.fr

## ABSTRACT

All-solid-state batteries hold great promise for electric vehicle applications due to their enhanced safety and higher energy density. However, further performance optimization requires a deeper understanding of their degradation mechanisms, particularly at the nanoscale. This study investigates the real-time degradation processes of an oxide-based all-solid-state micro-battery, using focused ion beam lamellae composed of LAGP as the solid electrolyte, $LiFePO_4$ (LFP) composite as the positive electrode, and $LiVPO_4$ (LVP) composite as the negative electrode. In situ electrochemical transmission electron microscopy (TEM) revealed critical degradation phenomena, including the formation of cracks along grain boundaries in the solid electrolyte due to lithium diffusion and mechanical stress. Additionally, the shrinkage of solid electrolyte particles and the formation of amorphous phases were observed. These findings highlight the importance of grain boundary dynamics and amorphization in the performance of solid electrolytes and provide insights into degradation mechanisms that can inform the design of more durable all-solid-state batteries.

**INTRODUCTION**

All-solid-state batteries (ASSB) are considered to be the future promising battery technology as they provide enhanced safety, higher energy density, simpler battery packing, and larger operable temperature range compared to conventional Li-ion battery (LiB) technology[1–5]. Despite the latest improvements in the solid electrolyte (SE) ionic conductivity and SSB performances, their widespread application is still limited due to various challenges like solid electrolyte-electrode interface[6–8], morphological transformations during the cycling process [9–11], dendrite growth and crack propagation[12–14] which triggers their failure mechanism. Getting a better insight into the degradation's origin is the key to develop high energy density ASSBs and ensuring their success in the energy storage market [15,16].

During the past several years, different *ex situ* studies[17,18] have been conducted to provide an understanding of the morphological, crystal structural changes or chemical evolution in the failure mechanism of all solid-state batteries. Unfortunately, in these studies, the battery is settled outside of the electrochemical system in a relaxation state. To provide a deeper comprehension of the degradation mechanism, a real-time monitoring of the dynamic processes is required during the electrochemical cycling through *in situ/operando* measurements [9,19,20]. In the past years, various characterization techniques such as Raman spectroscopy, scanning electron microscopy (SEM)[18], X-ray diffraction (XRD)[21], X-ray computed tomography (XCT)[22–24], X-ray photoelectron spectroscopy (XPS)[25] have been used to monitor the failure mechanism of solid-state batteries. Among the various characterization techniques, transmission electron microscopy (TEM) stands out for its exceptional ability to deliver morphological, structural, and compositional information with atomic-level spatial resolution [26–29].

In situ/operando TEM studies have significant potential to provide comprehensive and dynamic information at the nanoscale level [28,30,31]. Their application to solid-state batteries (SSBs) has however been challenging due to practical difficulties in fabricating a complete solid-state micro-battery, connecting to micro-devices and managing moisture-sensitive materials [28,30,31]. All-solid-state batteries utilizing oxide solid electrolytes offer significant advantages, including air stability and superior electron-beam stability, which simplify the microbattery fabrication process. In contrast, other solid electrolyte classes, such as sulfides and halides, exhibit high moisture sensitivity, necessitating handling in controlled atmosphere environments [32]. Cycling a full solid-state battery based on oxide solid electrolyte at room temperature (RT) without the presence of any liquid electrolyte, binder or additives is however challenging due to their low ionic conductivity and the presence of grain boundary reducing ionic diffusion [33,34]. Very few results have been obtained at RT on fully solid-state batteries employing oxide solid electrolytes without any liquid electrolyte additives to enhance their low ionic conductivity. This scarcity underscores the complexity of such solid systems and the significant improvements required [35,36]. For instance, E. Kobayashi *et al.*[36] showed the key concept with an all-solid-state phosphate battery using $Li_3V_2PO_4$ (LVP) as both positive and negative electrodes and $Li_{1.5}Al_{0.5}Ge_{1.5}(Po_4)_3$ (LAGP) as solid electrolyte obtaining a discharge capacity of 104 mAh.g$^{-1}$ at 80°C. Meanwhile, A. Aboulaich *et al.*[35] developed an LiFePO$_4$/LAGP/LVP all-solid-state battery achieving a discharge capacity of 80 mAh.g$^{-1}$ at 120 °C over 30 cycles.

In recent years, a limited number of *in situ* TEM studies have been published, employing two distinct configurations to elucidate various components of battery systems. In a first approach, Li-metal is deposited on an adjustable metal tip which induces polarization upon contact with the solid electrolyte allowing to monitor the (electro)chemical evolution at the Li metal/SE interface. M. McDowell *et al.*[11] used *in situ* electrochemical TEM to investigate the nanoscale reaction process of Li metal and LAGP particles and observed in real time a decrease in lithium particle volume, expansion of solid electrolyte particles indicating lithium diffusion inside and its amorphization. X. Liu *et al.*[37] utilized *in situ* TEM combined with electron energy loss spectroscopy (EELS) to investigate the LLZO solid electrolyte. Their observations revealed that lithium diffusion initiates at the grain boundaries (GB) during cycling rather than at the interfaces, leading to the segregation of isolated intergranular lithium.[37] In a second approach, a thin Focused Ion Beam (FIB) lamella comprising a fully functional micro-battery, encompassing a positive electrode, solid electrolyte (SE), and negative electrode, is integrated with a microchip to enable bias configuration. This approach is more complex as it provides comprehensive insights into the entire battery system, enabling the monitoring of all solid interfaces. However, this approach presents significant challenges, as it requires the fabrication of a micro-battery with dimensions in the range of several tens of micrometers, precise alignment with the microchip's contact points, and maintaining a thickness of 100 nm to ensure beam transparency. The seminal investigation by A. Brazier *et al.* [38] in 2008 represents the first *ex situ* cross-sectional observation of an all-solid-state nanobattery. This study compared pristine and cycled nanobatteries derived from micro-batteries, revealing the chemical migration of silicon and vanadium from the LVSO electrolyte into the SnO electrode. In 2011, Y. S. Meng *et al.* [39] prepared a FIB lamella of a solid-state battery, consisting of LVSO as the solid electrolyte, $LiCoO_2$ (LCO) as the cathode, and SnO as the anode. They observed significant damage to the electrolyte and electrodes during bias application, attributed to high local currents. Recently, C. Kubel *et al.* [40] succeed to realize the first *in situ* TEM study on a fluoride all-solid-state battery using Cu/C composite as cathode, $La_{0.9}Ba_{0.1}F_{2.9}$ as solid electrolyte and a nanocomposite of $MgF_2$, Mg, $La_{0.9}Ba_{0.1}F_{2.9}$ and C as anode. Post-cycling observations revealed the presence of voids and cracks, along with fluoride migration from the solid electrolyte into the cathode and copper diffusion into the solid electrolyte layer. Newly, a TEM study utilizing STEM-EELS on the $Li_7La_3Zr_2O_{12}$ solid electrolyte demonstrated that after bias application, the grain boundary regions exhibit a reduced bandgap, leading to increased electron density [37]. This facilitates the reduction of lithium metal, ultimately resulting in the formation of Li filaments that penetrate the solid electrolyte, causing a short circuit.

In our study, we successfully prepared a micro-battery FIB lamella, and we carried out an *in situ* electrochemical TEM experiment on an all-solid-state Li-ion battery FIB lamella containing $LiFePO_4$ as positive electrode, LAGP as the solid electrolyte and $Li_3V_2(PO_4)_3$ as the anode. An advanced approach was utilized to monitor the dynamic evolution of the morphology of all-solid-state batteries at the nanoscale. This methodology integrates Scanning Transmission Electron Microscopy-High-Angle Annular Dark Field (STEM-HAADF) imaging with several electron microscopy techniques, including Energy Dispersive X-ray (EDX) spectroscopy and Electron Energy Loss Spectroscopy (EELS), to determine the chemical changes. Additionally, 4D-STEM Automated Crystallographic Orientation Mapping (ACOM) analysis was employed to observe the alterations in crystal structure before and after the cycling process.

**RESULTS AND DISCUSSION**

A full stack all-solid-state battery can be obtained by various techniques such as atomic layer deposition[41], applied hot or cold pressure[42–44], 3D printing or spark plasma sintering (SPS)[35,45–47]. The SPS technique enables the rapid formation of high-quality interfaces between solid electrolytes (SE) and electrodes. This method significantly reduces the processing time while maintaining superior interfacial integrity, which is crucial for the performance of electrochemical systems. Herein, an all-solid-state battery composed of $Li_{1.5}Al_{0.5}Ge_{1.5}(PO_4)_3$ (LAGP) as the solid electrolyte, $LiFePO_4$ as the positive electrode and $Li_3V_2(PO_4)_3$ as the negative electrode (See experimental methods and **Figure S1** in the Supporting Information) was obtained in a single shot using SPS technique (See experimental methods) as observed in the schema displayed in **Figure 1a.** The electrodes are a composite of active material, solid electrolyte to ensure ionic conductivity in the electrodes and black carbon to provide the required electronic conductivity, with an optimum formulation of 6:3:1 (wt %) respectively. **Figure 1b-d** presents cross-sectional scanning electron microscopy (SEM) images of the all-solid-state battery, specifically designed for *in situ* TEM cycling applications. The LAGP solid electrolyte layer exhibits a thickness of only a few micrometers, enabling the extraction of a FIB lamella encompassing the entire battery within just a few tens of micrometers. This precise length is necessary to span the distance between the gold electrodes on the micro-chip, where the FIB lamella was subsequently connected. The solid electrolyte layer exhibits a variable thickness spanning a 2 - 50 µm range (**Figure S2**, Supporting Information), the thinner layer areas of SE allowing the successful extraction of a full micro-battery FIB lamella. The cross-section SEM-EDX analysis unveiled (**Figure S3**) a clear interface between the solid electrolyte and electrodes layers indicating no diffusion of the elements between battery components before the cycling process.

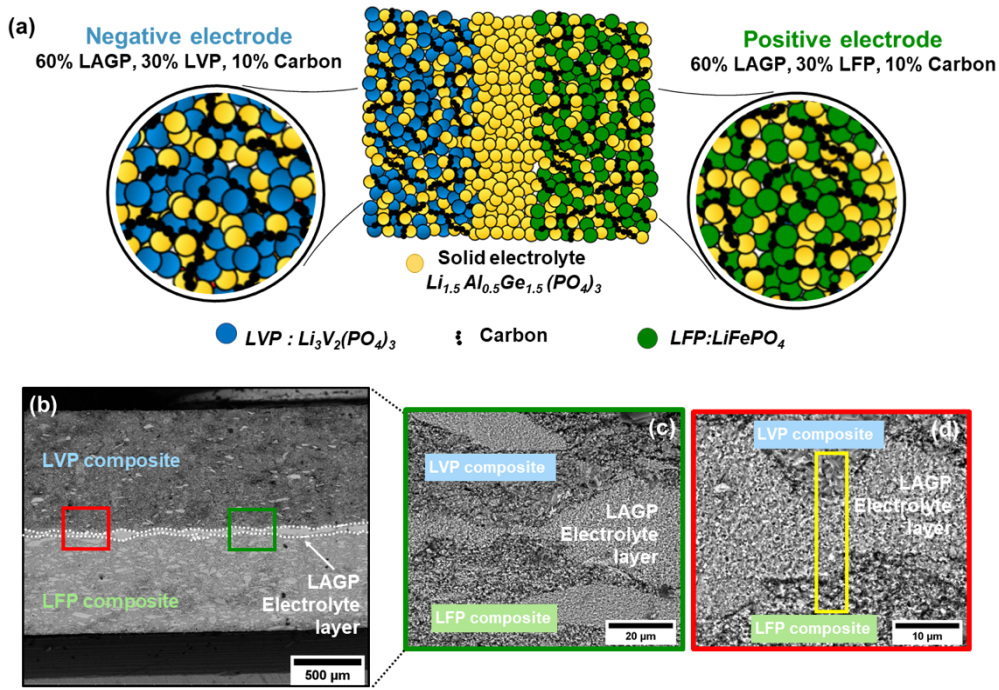

**Figure 1**. **Full stack all-solid-state battery based on LAGP solid electrolyte.** (a) Schematic of the LVP-LAGP-Carbon/LAGP/LFP-LAGP-Carbon all-solid-state battery prepared by Spark Plasma Sintering (SPS) showing the solid electrolyte (yellow) and the composite electrodes. (b-d) Back-scattering SEM cross section image of a solid-state battery prepared by SPS showing a very thin solid electrolyte layer and the two electrode composites. The yellow box in panel (d) highlights the region selected for FIB lamella preparation.

To conduct *in situ* electrochemical TEM experiments, it is essential to develop a micro-battery that meets several stringent criteria [48], including high electron-beam transparency and appropriate electrical contacts with the micro-chip. Achieving such a micro-battery is particularly challenging due to the necessity of avoiding Ga contamination during the preparation process. Additionally, establishing proper contact between the electrodes of the FIB lamella and the E-chip is crucial. Furthermore, the micro-battery must exhibit sustainable mechanical properties while being sufficiently thinned to ensure adequate e-beam transparency.[48] For this study, a FIB lamella was successfully extracted, from the bulk all-solid-state battery prepared by SPS, using a FIB technique for etching, micro-manipulating and contacting (See details in the Experimental methods). **Figure 2a** unveils our FIB lamella micro battery with a clear interface separation between the positive electrode (left), solid electrolyte and negative electrode (right) and a thickness of only few tens of micrometers for each layer. The electrodes of the FIB lamella cross section were connected to the electrodes of the E-chips (Protochips) (**Figure 2b,c**) using Gas Injection System (GIS) with Pt bridge deposited and irradiated (e-beam) (**Figure 2d**) to ensure good conductivity between the micro-battery and E-chip following optimized conditions from previous work[28]. To prevent short-circuiting in the micro-battery, the platinum layer situated between the positive and negative electrodes was meticulously milled. Subsequently, to ensure beam transparency, a final thinning process was performed, reducing the thickness down to 100 nm at the interface between the solid electrolyte and the positive electrode. **Figure 2e** displays our ready to be cycled all solid-state micro-battery containing LAGP as the solid electrolyte and LVP and LFP composites as the negative and positive electrodes, respectively. To ensure enough mechanical stability, the thin and beam transparency area was ensured at the interface of the solid electrolyte with the LFP cathode.

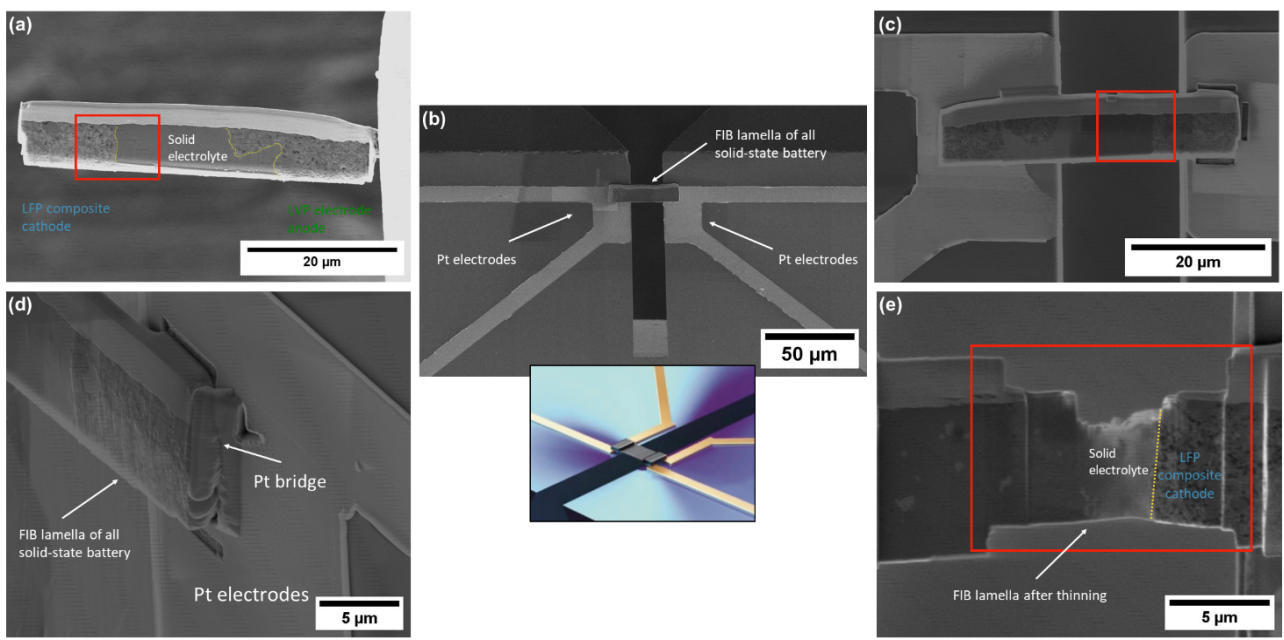

**Figure 2**: **Micro-battery preparation using a FIB lamella connected to a microchip via a Pt bridge contact.** (a) Schematic illustration showing (a) the Protochips Fusion holder and the E-chip used for the *in situ* experiment (b) FIB lamella micro-battery containing solid electrolyte and the electrodes in < 50 µm extracted from a larger battery prepared by SPS (c) FIB lamella micro-battery connected on the Protochips E-chips (d) FIB lamella connected on the E-chips using platinum connections (e) Removal of the top Platinum layer between the electrodes and thinning down the FIB lamella for e-beam transparency.

To obtain structural information with spatial resolution about the solid electrolyte, electrode and the interface before the cycling process, 4D-STEM ACOM (Automated Crystal Orientation Map) analysis was performed on the pristine micro-battery. The 4D-STEM method[49–51] uses a scanning nano-probe approach and precession beam (angle at 0.7°) to acquire electron diffraction patterns (**Figure 3h**) over an analyzed area. The collected electron diffraction patterns are further automatically indexed using a template matching method (Astar-NanoMegas) that generates banks of simulated diffraction patterns, based on CIF files of all expected phases, with all possible orientations. To avoid electron beam damage on the micro-battery, the scan time was limited to 20 ms per electron diffraction. **Figure 3a** reveals the index map of the analyzed area, corresponding to the interface between the solid electrolyte - negative electrode (analysis performed on another FIB lamella extracted from the same bulk battery). The orientation map along the Z axis (**Figure 3b**) shows no preferential orientation for the $Li_3V_2(PO_4)_3$ phase, but for the solid electrolyte the [100] orientation was the most encountered one, which can be explained by the uniaxial compression SPS technique. The grain boundary (GB) map presented in **Figure 3c** exhibits in black the grain boundaries between the same phase and in green the grain boundaries between different phases. A higher amount of GB is presented in the composite electrode compared to the solid electrolyte possibly indicating a higher ionic conductivity in the separator layer compared to the composite electrode. In solid electrolytes, the grain boundaries are considered as the initial source of cracks due a low fracture toughness[52] and their increase in electron density that facilitates the lithium-ion reduction to lithium metal[37]. The reconstructed phase map (**Figure 3d**) reveals in pink the active material (7.9 %), the solid electrolyte in yellow (68.8%), and the additive carbon in blue (23.5%). The acquired electron diffraction patterns for the LAGP, LVP and additive carbon displayed in the **Figure 3e-g**, in which the scattered ring corresponds to amorphous carbon contribution. Based on the reconstructed map, it is evident that the solid electrolyte is highly integrated among the LFP grains, thereby ensuring optimal ionic conductivity within the electrode. This meticulous incorporation significantly enhances the electrochemical performance, indicating a robust and efficient electrode architecture. Based on the phase map, a heterogeneous distribution of the additive carbon in the negative composite electrode can be speculated as a higher concentration than expected (10 wt%) was found in the analyzed area. Moreover, a slight migration of carbon and active material from the electrode composite layer in the SE layer at the grain boundaries was observed, the phenomena probably occurred during battery assembly or the sintering process. The presence of carbon in the separator layer can result in an electronic conductivity increase which may promote dendrite formation[53].

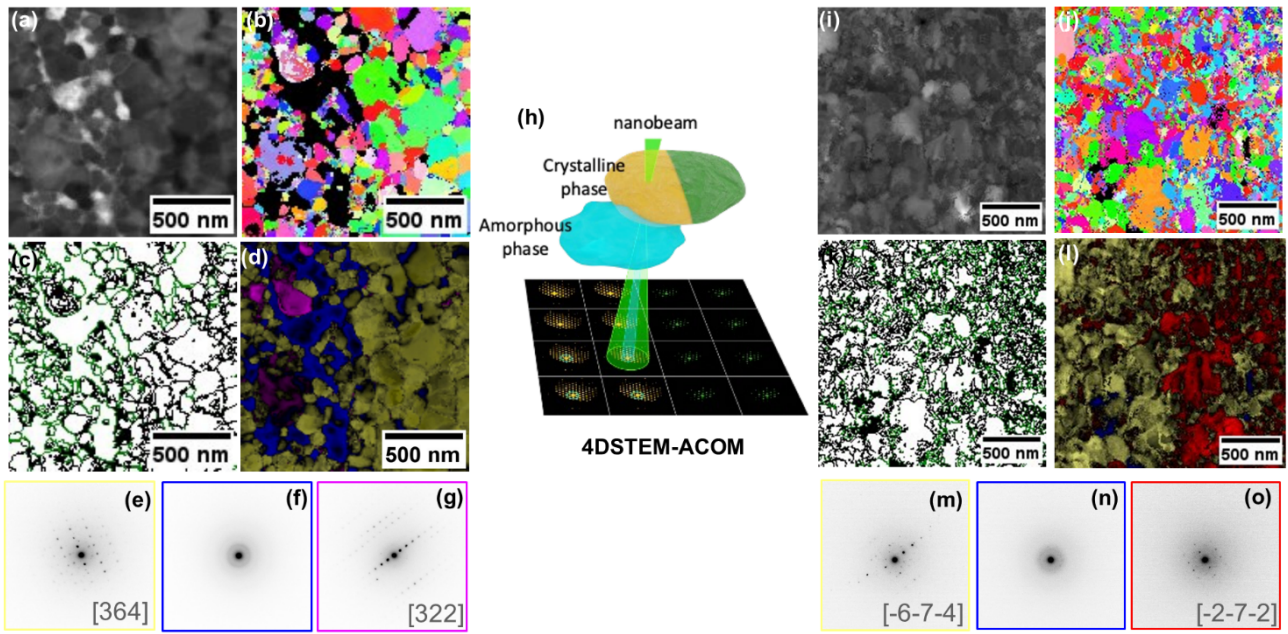

**Figure 3**: **Structural mapping analysis of pristine cathode and anode electrodes using 4D-STEM ACOM technique.** 4D-STEM ACOM analysis performed on the negative electrode ($Li_3V_2PO_4$) /solid electrolyte interface showing (a) the bright field map of the analyzed area (b) orientation map according to X direction (c) grain boundaries map (d) reconstructed phase map based on the acquired electron diffraction patterns of the (e) solid electrolyte (yellow), (f) carbon (blue) and (g) $Li_3V_2PO_4$ (purple). (h) Schematic of the 4D-STEM analysis principle and its result performed on the $LiFePO_4$/solid electrolyte interface after the electrochemical bias showing (i) index map (matching score between experimental and simulated ED from known resolved structure) corresponding to the analyzed area (j) orientation map according to Z direction (k) grain boundaries map (l) reconstructed phase map based on the matching pattern between the simulated electron diffraction pattern and acquired ones. Examples of acquired electron diffraction patterns of (m) solid electrolyte (yellow) (n) carbon (blue) and (o) $LiFePO_4$ (red).

The STEM-HAADF (Scanning Transmission Electron Microscopy - High Angle Annular Dark Field) image of the ASSB micro-battery before bias application is displayed in **Figure 4a** where on the right side the LFP composite electrode can be observed and on the left side the solid electrolyte with variable thickness. To conduct *in situ* cycling experiment of FIB lamella micro-battery was mounted to a Fusion-Select holder from Protochips, as illustrated in **Figure 4c**. This configuration allows the application of an electrical bias. Electrochemical measurements were executed using a Keithley system, which controlled the potential applied to the micro-battery. The cyclic voltammetry (CV) curve of the LVP/LAGP/LFP micro-battery can be observed in **Figure 4d**. A CV sweeping rate of 10 mV/s was applied on the micro-battery in a range of potential between 0 and 5 V. The cyclic voltammetry curve reveals that our micro-battery exhibits substantial resistance, which induces high polarization and affects its electrochemical performance. The observed pronounced polarization is likely attributable to several factors. Primarily, it may arise from the suboptimal interfacial contact area between the electrode components. Additionally, the intrinsic resistances within the all-solid-state battery play a significant role. This is further exacerbated by the presence of numerous grain boundaries, which introduce additional resistance[52,54]. Finally, the resistance from the holder and the micro-chip also contributes to the overall polarization.

Conversely, during the electrochemical reaction, it was observed that the electron beam exerted an influence on the electrochemical curve, as illustrated in blue in **Figure 4d**. This interaction underscores a significant effect of the electron beam on the electrochemical dynamics under investigation. To minimize potential interference during the *in situ* experiment, the electron beam was deactivated as much as possible, except during the acquisition of STEM images.

Although the CV curve obtained during the *in situ* experiment exhibits no clear presence of electrochemical activity. In an *in situ* TEM experiment utilizing an electrochemical setup, significant electrical resistance within the device, especially at the electrolyte-electrode interface, can induce considerable polarization. This phenomenon results in a larger voltage drop across the interface than predicted by the applied potential alone. Consequently, the actual potential experienced by the electrode diverges significantly from the applied potential, distorting or even completely obscuring electrochemical responses in the cyclic voltammogram curve. Excessive polarization can indeed hinder observable electrochemical reactions within the applied potential range during CV measurements. The measured current remains minimal because the substantial overpotential caused by high polarization is insufficient to drive the desired electrochemical processes. Therefore, a CV curve that shows no clear electrochemical reactions during an *in situ* electrochemical experiment likely indicates excessive polarization within the setup which is the case in this in situ experiment. To address this issue, optimizing experimental parameters and minimizing the device's overall resistance is crucial. These adjustments will ensure proper charge transfer, enabling the observation of the intended electrochemical phenomena, which is a strong perspective for our next *in situ* TEM experiment to improve the reliability of the electrochemical reactions.

From the initial state of the micro battery (**Figure 4a**) to the final state of the battery (**Figure 4b**) various phenomena that indicated that an electrochemical reaction was conducted were observed (see also **video 1** in the Supporting Information). The solid electrolyte particles were observed to undergo morphological changes during bias applications, **Figure 4e** indicates in yellow reveals that most of the LAGP particles shows a shrinkage in the surface particle area together with an important increase in the grain boundaries region. This observation indicates that the substantial disparity in lithium conductivity between the bulk particle and the grain boundary region may catalyze the initiation of electrochemical reactions at the particle/grain boundary interfaces[33]. Grain boundaries are known to have distinct physical and chemical properties compared to the bulk material, often exhibiting higher defect concentrations, altered electronic structures, and modified ionic conductivities. In the context of Li-ion conductors, the grain boundary regions frequently present a higher energy barrier for lithium-ion transport compared to the bulk material, resulting in significantly lower ionic conductivity[55]. This conductivity gradient can lead to localized concentration gradients of lithium ions at the interfaces between the bulk particles and the grain boundaries. Such gradients can drive electrochemical reactions by creating regions of high electrochemical potential differences, effectively acting as hotspots for reaction initiation. For instance, in solid-state batteries, these interfacial regions can become active sites for lithium plating and stripping, influencing the overall performance and stability of the battery. Additionally, the electrochemical reactions at these interfaces can contribute to phenomena such as dendrite formation, which is a critical challenge for the safety and longevity of lithium-based batteries[56].

During the electrochemical reaction, the most notable change observed was the development of cracks within the solid electrolyte layer. These cracks originated exclusively in the solid electrolyte layer, with no signs of initiation in the electrode. Furthermore, it was evident that these cracks expanded along the grain boundaries and therefore between the main LAGP grains. This phenomenon can be attributed to lithium migration into the grain boundaries, leading to increased stress levels affecting the solid electrolyte's mechanical characteristics. This observation supports the previously observed phenomena highlighting the grain boundaries as a critical site for structural and electronic weaknesses[37,52]. As a result, cracks emerge due to the heightened tension caused by the diffusion of Li in the grain boundaries. Additionally, the inherent brittleness of the oxide solid electrolyte may further facilitate the propagation of these cracks. Crack formation was observed in both the thick (**Figure 4f**) and thin regions (**Figure 4g**) of the FIB lamella, likely induced by mechanical stress associated with electrochemical cycling and state of charge. This suggests that the experimental conditions may have exceeded the stability window of LAGP, in line with the high polarization observed in our *in situ* setup.

The STEM-HAADF images displayed in the **Figure 4h** indicate a region in the solid electrolyte where the presence of a few darker secondary particles can be observed, this could correspond to a compound with a higher amount of lithium or a phase containing lighter elements than the solid electrolyte. Interestingly enough, during the cycling process, in these particles a change of contrast from darker to lighter was observed, right before inducing the appearance of cracks in the thin part of the solid electrolyte in this region. This modification in contrast could be associated with possible lithium diffusion/consumption towards the electrodes side resulting in mechanical stress in the solid electrolyte. Furthermore, another assumption explaining this contrast feature could be related to a conductivity gradient between the bulk particles and the grain boundaries leading to localized Li-ion concentration gradients, which can induce extra electrochemical reaction by creating regions of high overpotential differences. The cracks in the thin part of the micro-battery were observed to initiate from an area containing secondary phases, which undergo changes during cycling process suggesting that the two phenomena could be linked.

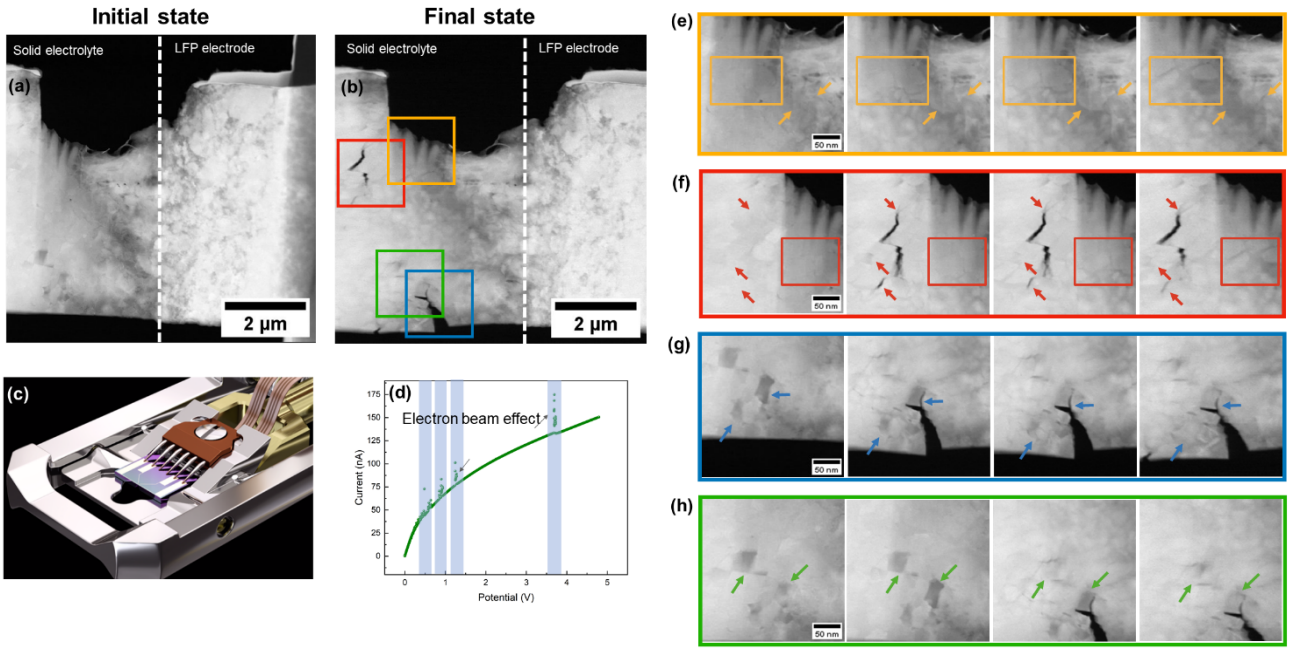

**Figure 4**. *In situ* **electrochemical TEM experiment of all-solid-state micro-battery.** ASSB FIB lamella (a) before cycling showing the positive electrode /solid electrolyte interface and (b) after *in situ* STEM cycling process. (c) Image representation of the Fusion-Select holder used to perform the electrochemical reaction and (d) the electrochemical curve corresponding to the cycling of the ASSB battery. (e-h) Dynamic evolution of the ASSB during the cycling process showing (e) particles shrinkage and grain boundaries increasing (f-g) the presence of cracks induces by the bias in different parts of the solid electrolyte and (h) contrast change during the electrochemical reaction.

To understand better the structural changes occurring in our micro-battery after the bias application, a post-mortem 4D-STEM ACOM analysis was carried out. During the 4D-STEM ACOM analysis[49,51,57], the electron diffraction patterns (**Figures 5a-e**) were scanned over an area in the positive electrode as shown in the index map displayed in **Figure 5f**. A phase map was reconstructed based on the acquired electron diffraction patterns. **Figure 5g** shows an area of the positive electrode in which the active material is displayed in purple, the solid electrolyte in yellow, the carbon in blue, and the amorphous region in black. The distinction between the carbon and the amorphous region was made using the presence of the intense amorphous ring in the electron diffraction pattern for black carbon (**Figure 5b-c**). The amorphous region (LAGP solid electrolyte) observed to appear after the cycling process with the 4D-STEM analysis is found in a 15.4 % of the total analyzed area. A similar amorphization phenomena was noticed by McDowell *et al.* using *in situ* TEM where the amorphization of LAGP at the interface with lithium metal was detected the after bias application and confirmed by XRD and Raman spectroscopy on Li/LAGP/Li cells[11]. The amorphization of LATP solid electrolyte in contact with Li-metal after one hour of reaction was reported by Cheng *et al.*[58] showing that the negative voltage initiates the amorphization transformation. Moreover, Zhu *et al.*[59] demonstrated that the electrochemical reaction, driving the lithiation process, induces the amorphization of LATP. The driving force of this amorphization process is the LAGP degradation, which was identified by EELS spectroscopy as shown in Figure 6d), either by Li-ion loss or by agglomeration of lithium inside the solid electrolyte part. The phenomenon of amorphization has been extensively documented in numerous studies on all-solid-state batteries[60]. This phenomenon appears to be strongly associated with deviations from the chemical and electrochemical stability of solid electrolytes, as well as with the chemo-mechanical challenges commonly encountered in these systems[61]. In multilayered thin-film structures, solid-state amorphization through diffusion-driven reactions has been noted, particularly when the film thickness approximates the interface width[62]. This process results in the generation of an amorphous phase, forming from a thermodynamically favored crystalline intermediate compound failing to precipitate. Notably, the emergent glassy state exhibits reduced free energy compared to the unreacted constituents, rendering it metastable in nature. Sun *et al.*[63] demonstrated the electrochemically induced crystalline-to-amorphous transformation in sodium-ion solid electrolytes, emphasizing the crucial role of kinetics in the formation of the amorphous phase because the molecular dynamics calculations revealed that crystalline phases are thermodynamically favored.

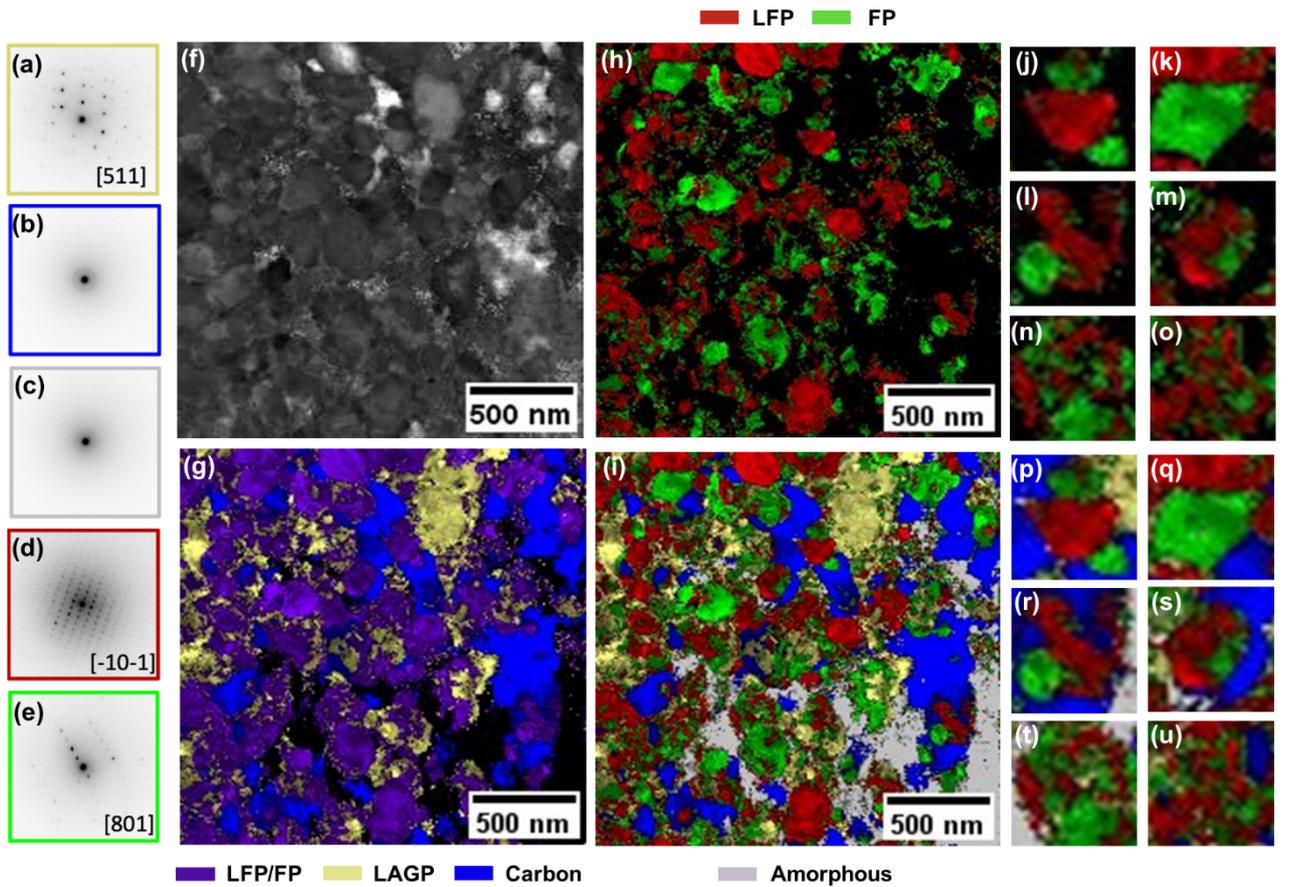

**Figure 5. 4D-STEM structural maps revealing local electrochemical activities.** 4D-STEM ACOM analysis performed on the LFP composite after the electrochemical bias showing electron diffractions of (a) LAGP (b) carbon, (c) amorphous area (d) LFP and (e) FP phase. (f) The reconstructed Index map shows the LAGP/LFP interface after cycling and the different phase maps show: (g) the positive electrode where the active material is represented in purple, solid electrolyte with yellow, carbon with blue, and amorphous areas with black (h) phase map that focus on the particles of the active material (LFP and FP) where in red are represented the LFP particles and in green the FP particles for better visualization of the LFP/FP interfaces and (i) overall phase map where LAGP (20.5 %) is represented in yellow, LFP (24.5 %) in red, carbon (18.5%) in blue, FP (21 %) represented in green and (15.4 %) amorphous points represented with gray. (i-o) Zoom in focusing on the interfaces of the LFP/FP phase map and (p-u) zoom in on the same area phase map where all the elements are presented.

LiFePO$_4$ de-lithiation process after the electrochemical reaction was probed using 4D-STEM mapping analysis. Both LiFePO$_4$ and FePO$_4$ crystals present olivine structure (Pnma space group) with a slight difference of 5% on the *a* lattice parameters, 3,7% on the *b,* and 1.9% on the *c* and despite them presenting such close values in the lattice parameters, and thus making it impossible to distinguish them by X-ray diffraction, 4D-STEM analysis using precession reports a clear separation between the two phases [64]. **Figure 5h** exhibits the phase map of the active materials after the bias application where the LiFePO$_4$ phase is displayed in red, the FePO$_4$ (FP) phase in green, and the other phases in black (carbon, amorphous and SE). The active material phase mapping confirms the presence of FePO$_4$ structure, being consistent with the delithiation process occurring over the electrochemical cycling.

Lithiation dynamics in LiFePO$_4$ materials are described by several models, each providing different insights into how lithium ions are inserted and extracted during battery operation. For instance, the shrinking core model suggests that lithiation begins at the surface and progresses inward, with a distinct boundary separating the lithiated and delithiated phases. However, this model struggles to explain the high-rate performance observed in nanoscale LiFePO$_4$, in which phase boundaries are less distinct. The mosaic model proposes that multiple nucleation points exist within a particle, leading to a patchwork of lithiated and delithiated regions, which better accounts for dynamic lithium diffusion in smaller particles. The radial core-shell model focuses on radial diffusion, with a lithiated outer shell surrounding a delithiated core, but like the shrinking core model, it is limited in explaining nanoparticle behavior. The domino-cascade model suggests that entire particles are either fully lithiated or fully delithiated, and this mechanism is particularly relevant for larger particles, where phase transitions occur more uniformly across the entire particle.

At the nanoscale, particles also show a reduction in the phase separation behavior due to the influence of surface effects, and studies have shown that at high current densities, solid-solution mechanisms may dominate. These findings highlight that the lithiation dynamics in LiFePO$_4$ are influenced by multiple factors, including particle size, crystal orientation, and defect structures. Understanding and controlling these factors is essential for optimizing the performance of LiFePO$_4$-based lithium-ion batteries.

A closer look at the phase map can help us to identify the lithiation mechanism of FePO$_4$/delithiation mechanism of LiFePO$_4$ at the nanoscale. The active material phase map (**Figure 5h**) shows that the large particles are mostly fully lithiated or delithiated confirming the presence of the domino-cascade process model, while the small particles are more a mix between lithiated and delithiated particles following a mosaic model. The particles that are fully lithiated or delithiated present the same orientation while the particles that present a mix between lithiated and delithiated phase shows also a combination of orientations suggesting that the orientation of the crystals plays a role in the lithiation/delithiation mechanism. The overall phase map with all the phases is presented in **Figure 5i**. On a closer look we can observe that the presence of mixed lithiated/delithiated phases is mostly located around the amorphous area that appeared after the electrochemical reaction. The proximity between the amorphous regions and the highly mixed phases indicates significant electrochemical activity, likely linked to the amorphization phenomenon caused by the instability of the LAGP. This suggests that the structural degradation in these areas plays a key role in driving electrochemical processes.

To evaluate the chemical evolution before and after the electrochemical reaction, Electron Energy Loss Spectroscopy (EELS) experiments were performed. Figure 6a presents the EELS spectra of the positive electrode, where the spectrum prior to bias application is shown in blue, and the spectrum after the delithiation process is shown in red. The post-reaction spectrum exhibits a distinct peak at around 5 eV, which strongly suggests successful delithiation of $LiFePO_4$ during the *in situ* experiment. This peak is associated with the valence spectrum of $FePO_4$. As noted in the literature, the 5 eV peak is characteristic for the $FePO_4$ phase and is absent in the $LiFePO_4$ phase. This distinction is due to differences in the electronic structure between the two phases. Specifically, the presence of the peak in $FePO_4$ is linked to the empty states in the Fe 3d orbitals, which are not present in the $LiFePO_4$ phase, in which these states are filled due to lithium insertion. Thus, the peak at 5 eV serves as a useful marker in EELS spectra to identify fully delithiated particles in the transition process. Therefore, the clear emergence of this feature in the post-delithiation spectrum confirms the transition from $LiFePO_4$ to $FePO_4$, thereby verifying the electrochemical reaction progression. Furthermore, the absence of this peak before the electrochemical process emphasizes the stability of the $LiFePO_4$ phase in the initial state. In addition, a clear decrease of the lithium K-edge located at 60 eV after the electrochemical process indicates a modification in the lithiation composition inside the iron phosphate crystals.

**Figure 6d** shows EELS spectra of LAGP solid electrolyte before and after delithiation exhibiting a significant modification of low loss bands associated to LAGP materials with a clear change in Li K-edge peak at 60 eV. This modification of LAGP solid electrolyte after electrochemical reaction is related to the amorphization phenomena observed in 4D-STEM structural maps.

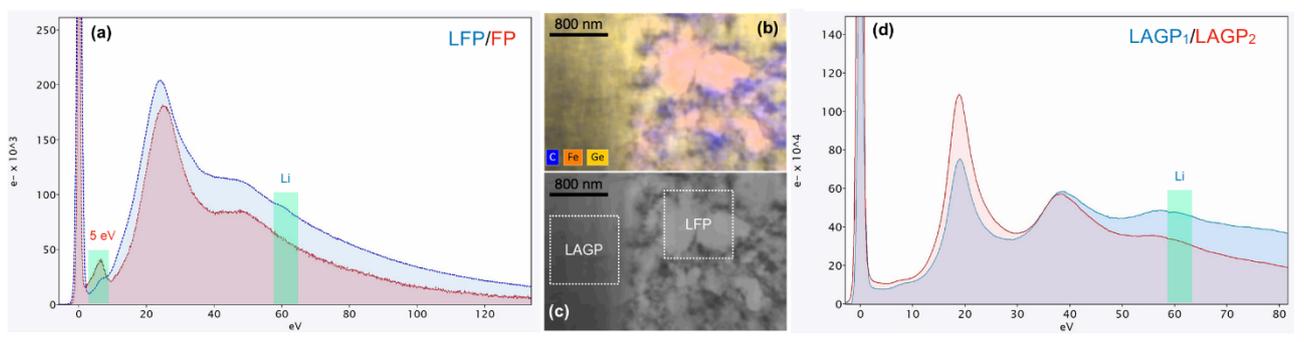

**Figure 6. EELS spectroscopy of cathode active and solid electrolyte materials.** (a) EELS spectra of LFP particles inside positive electrode before and after delithiation with bond a 5 eV corresponding to the presence of FePO4 and a modification of Li K-edge peak at 60 eV after lithiation. (b,c) EDX map and STEM-HAADF images of LAGP/cathode interface with probed zone area. (d) EELS spectra in low loss energy of LAGP solid electrolyte before and after delithiation exhibiting slight modifications with presence of Li K-edge peak at 60 eV.

During the electrochemical cycling of ASS battery, the formation of amorphous phases in solid electrolytes is a critical phenomenon that has been observed and studied extensively. This transformation from crystalline to amorphous structures is often linked to mechanical stresses, interfacial reactions, and the instability of certain electrolyte materials under cycling conditions. For example, in sulfide-based electrolytes like $Li_7P_3S_{11}$, amorphization occurs due to lithium-ion movement and interfacial instability, which can degrade ionic conductivity over time. This issue has also been observed in materials like LLZO (lithium lanthanum zirconium oxide) and in polymer-based electrolytes where phase changes impact the long-term stability of the battery. Amorphous phases can form because of significant structural disorder during cycling, particularly at the electrode-electrolyte interface. This can cause phase transformations and a breakdown in the crystalline order, reducing the electrolyte's ability to effectively conduct lithium ions. While some amorphous phases can still provide adequate ionic conductivity, they generally exhibit poorer stability and lower performance compared to their crystalline counterparts. In certain systems, however, amorphous materials have been deliberately engineered to enhance flexibility and accommodate volume changes during cycling, improving battery performance and durability.

**CONCLUSIONS**

All-solid-state micro-battery FIB lamellas composed of an oxide solid electrolyte, LAGP, a composite positive electrode based on LFP and a composite negative electrode based on LVP were successfully prepared for *in situ* electrochemical TEM application. FIB lamellas micro-batteries were extracted from a larger all-solid-state battery specially designed with a controlled thin layer of SE, prepared by SPS, allowing for obtaining a densified battery with good contact interfaces connecction between the battery components.

Electric bias was applied to the micro-battery and its dynamic evolution was monitored at a nanoscale level using advanced electron microscopy techniques. During *in situ* TEM cycling, although the CV curve showed resistance, various dynamic processes were observed on the micro-battery. STEM-HAADF images showed that SE particles are shrinking during bias application meanwhile the grain boundary of SE particles region is increasing. Other observed phenomena were the apparition of cracks in the solid electrolyte, the cracks were observed to initiate at the grain boundaries region. Secondary particles spotted in the separator layer were also observed to undergo contrast, right before cracks apparition, one assumption for this phenomena being lithium diffusion towards the electrode side induced mechanical stress and enhanced crack propagation. The brittleness of the oxide solid electrolyte more likely contributed to the enhancement of the crack propagation, phenomena observed only in the separator layer.

The 4D-STEM ACOM analysis revealed the apparition of an amorphous phase after the bias application, which was associated with the amorphization of the solid electrolyte. Another important finding in this study was the observation of the delithiation mechanism of $LiFePO_4$ at the nanoscale. The 4D-STEM structural map showed that the larger $LiFePO_4$ particles that present the same orientation follow the domino-cascade delithiation process, meanwhile more of a mix of lithiated/deliathiated particles, for smaller scale, was spotted around the new formed amorphous phase. STEM-EELS analysis performed before and after the bias application confirms the electrochemical reaction in the system and the presence of $FePO_4$ phase as well as the strong modification of LAGP solid electrolyte.

Based on the *in situ* electrochemical TEM experiment performed on the oxide solid-state battery coupled with advanced characterization techniques, we confirm as a degradation mechanism the shrinkage of solid electrolyte particles with an increase in the grain boundary area, the apparition of cracks at the grain boundaries area due to Li movement, and the amorphization of the solid electrolyte.

**EXPERIMENTAL METHODS**

*Synthesis of the $Li_3V_2(PO_4)_3$*

$Li_3V_2(PO_4)_3$ as anode material was synthetized using a solid-state reaction where the precursors $VPO_4$ and $Li_3PO_4$ (2:1 molar ratio) were mixed using a mortar and pestle. The obtained composition was placed in a furnace oven in an alumina crucible under an argon atmosphere at a temperature of 800 °C for 2h. The XRD profile fitting analysis of the $Li_3V_2(PO_4)_3$ powder is displayed in the Supporting information, **Figure S1**.

*All-solid-state battery formulation*

The positive and negative cathode composites were formulated from a mixture containing a weight ratio of 10:60:30 of carbon black (C45): solid electrolyte (LAGP) and active material (AM), $Li_3V_2(PO_4)_3$ in the case of the negative electrode and $LiFePO_4$ in the case of the positive electrode. Each electrode powder was mixed for 30 minutes using a Mixer Mill 8000M (SPEX).

*Spark Plasma Sintering*

An FCT GmbH HPD10 Spark Plasma Sintering with the chamber in a glovebox (controlled atmosphere) was used to obtain the all-solid-state Li-ion battery in one single shot. During the sintering program, a pressure of 100 MPa was initially applied on the all-solid-state battery, and then a heating program with a rate of 1 °C/s until 680 °C followed by a plateau of 5 minutes at the maximum temperature. A fast cooling until room temperature was performed in a controlled period of time (5 minutes). The solid-state battery was polished using sandpaper to remove the graphite paper.

*FIB lamella preparation and lamella connection to the E-chip*

The micro-battery was prepared using a Focused Ion Beam (FIB) FEI STRATA DB 235 microscope. Local platinum deposition assisted by the ion beam is used to protect the microbattery surface (no platinum is deposited with the electron beam). Initial milling to obtain a cross-sectional FIB lamella containing all 3 layers of the battery is done using a current of 7 nA. FIB lamella is transferred from the sample to Protochips E-chip by using a Kleindiek micromanipulator. FIB lamella cross section is connected to electrodes Protochips chip using a platinum deposition on both sides of the microbattery electrodes. The initial platinum layer used for protection of the microbattery is slowly milled in order to avoid a short-circuit in the battery as the two electrodes are connected between them by the platinum layer. A final thinning down to 100 nm in the FIB lamella for the electron beam transparency was realized with a current of 100 pA.

*4D-STEM ACOM data acquisition and data analysis*

The 4D-STEM Astar system was employed for automatic TEM phase-orientation mapping using a precession electron beam frequency of 100 Hz. Electron diffraction patterns of the sample were collected at an electron energy of 200 kV with a precession angle of 0.7°. Diffraction patterns, acquired using a CMOS Oneview camera, were realigned via ASTAR software. Phase identification was performed through pattern matching, using CIF files corresponding to the expected phases and reaction products. The ePattern software [51] was used to improve the 4D-STEM dataset for the pattern-matching process.

*In situ TEM cycling*

Thinned FIB lamella connected on the Protochips E-chip was mounted on the Fusion Select holder (Protochips) and further inserted in the TEM column. A Keithley system was used to drive the electrochemical measurements controlling the potential applied to our FIB lamella. Various scan rates from 1mV/s up to 10mV/s were used to drive the electrochemical reaction between 0 and 5 V. A Tecnai F20 S-Twin operating at 200kV was used for the acquisition of the STEM images.

**SUPPORTING INFORMATION**

The supporting information containing X-ray diffractogram and CV profile of Li3V2PO4 synthesis, solid electrolyte thickness in the bulk battery from where the micro battery was extracted, STEM-EDX analysis of the micro battery before and after *in situ* cycling is provided free of charge.

**ACKNOWLEDGEMENTS**

S.C. and A.D. gratefully acknowledge the financial support from the Hauts-de-France region and Protochips for providing the instrumentation used in the experiments. S.C. and M.D. also acknowledge the financial support from Nanyang Technological University. The authors express their gratitude to Protochips for providing the TEM holder for demonstration purposes, which enabled the experiments, and for their valuable contributions. All authors thank the UPJV and RS2E electron microscopy platforms for granting access to their facilities.

**AUTHORS CONTRIBUTIONS**

Conceptualization, A.D.; methodology, A.D. and S.C; software, N.F. and J.J. ; validation, A.D. and S.C ; formal analysis, A.D., N.F. ; FIB lamellas, D.T. and S.C ; in situ TEM experiments, S.C., N.K. and R.S. and N.F.; writing original draft preparation, S.C and A.D. ; writing review and editing, A.D. and A.J. ; supervision, A.D.; project administration, A.D., S.A. and M.D. ; funding acquisition, A.D. and M.D. All authors participated in the discussion and revision of this paper and finally approved this work.

**COMPETING INTERESTS**

The authors declare no competing financial or non-financial interests.